\documentclass[showpacs, oneside, twocolumn, amsmath, amssymb,nofootinbib,superscriptaddress]{revtex4-1}

\usepackage{cases}
\usepackage{amsfonts}
\usepackage{dcolumn}
\usepackage{bm}
\usepackage{bbm}
\usepackage{graphicx}
\usepackage{xcolor}
\usepackage{array}
\usepackage{subfigure}
\usepackage{hyperref}
\usepackage{cancel}
\usepackage{empheq}

\newcommand{\be}{\begin{equation}}
\newcommand{\ee}{\end{equation}}
\newcommand{\ba}{\begin{eqnarray}}
\newcommand{\ea}{\end{eqnarray}}

\newcommand{\gsim}{\mathrel{\hbox{\rlap{\lower.55ex \hbox {$\sim$}}
                   \kern-.3em \raise.4ex \hbox{$>$}}}}
\newcommand{\lsim}{\mathrel{\hbox{\rlap{\lower.55ex \hbox {$\sim$}}
                   \kern-.3em \raise.4ex \hbox{$<$}}}}

\renewcommand{\v}[1]{\ensuremath{\mathbf{#1}}} % for vectors

\renewcommand\O{\mathcal{O}}
\newcommand\R{\mathcal{R}}
\newcommand\s{\mathcal{S}}
\newcommand\A{\mathcal{A}}
\newcommand\cc{\mathcal{C}}
\newcommand\dd{\mathcal{D}}

\newcommand{\ac}[1]{\textcolor{red}{(AA: #1)}}

\def\equationautorefname~#1\null{Eq.~#1\null}

\begin{document}
\centerline{DESY-19-119}

\title{Orbital Inflation: inflating along an angular isometry of field space} %where do the symmetries hide?}

\author{Ana Ach\'{u}carro}
\email{achucar@lorentz.leidenuniv.nl}
 %\altaffiliation[Also at ]{XYZ University.}%Lines break automatically or can be forced with \\
 \affiliation{%
 Institute Lorentz of Theoretical Physics, Leiden University, 2333 CA Leiden, The Netherlands}
 \affiliation{Department of Theoretical Physics, University of the Basque Country, UPV-EHU 48080 Bilbao, Spain}
\author{Yvette Welling}%
 \email{yvette.welling@desy.de}
\affiliation{%
 Leiden Observatory,
 Leiden University, 2300 RA Leiden,
 The Netherlands}
 \affiliation{%
 Deutsches Elektronen-Synchrotron DESY, Notkestra{\ss}e 85, 22607 Hamburg, Germany}

\date{\today}

\begin{abstract}
The Cosmic Microwave Background (CMB) gives us a glimpse of the primordial perturbations. Their simplicity, so well described by single-field inflation,  raises the question whether there might be an equally simple multi-field realization consistent with the observations.
We explore the idea that an approximate 'angular' shift symmetry in field space (an isometry) protects the dynamics of coupled inflationary perturbations. This idea relates to the recent observation that multi-field inflation mimics the predictions of single-field inflation, if the inflaton is efficiently and constantly coupled to a second massless degree of freedom (the isocurvature perturbation) \cite{Achucarro:2019pux, Achucarro:2016fby}.
In multi-field inflation, the inflationary trajectory is in general \textit{not} aligned with the gradient of the potential. As a corollary the potential does not reflect the symmetries of perturbations.  
We propose a new method to reconstruct simultaneously a two-field action and an inflationary trajectory which proceeds along an `angular' direction of field space, with a constant radius of curvature,
and that has a controlled mass of `radial' isocurvature perturbations (entropy mass). We dub this `Orbital Inflation'. In this set-up the Hubble parameter determines the behavior of both the background and the perturbations. %In particular, a shift symmetry in the radial direction induces a zero entropy mass. Furthermore, 
First, Orbital Inflation provides a playground for quasi-single field inflation \cite{Chen:2009zp, Chen:2009we} because the couplings between perturbations are controlled and constant on the trajectory, up to slow roll corrections. Second, the exquisite analytical control of these models allows us to exactly solve the phenomenology of Orbital Inflation with a small entropy mass and a small radius of curvature, a regime not previously explored. The predictions are single-field-like, although the consistency relations are violated. Moreover, the value of the entropy mass dictates how the inflationary predictions fan out in the ($n_s$, $r$) plane. Depending on the size of the self interactions of the isocurvature perturbations, the non-Gaussianity parameter $f_{NL}$ can range from slow-roll suppressed to $\O(\text{a few})$. %Finally, larger deviations from gaussianity are possible and can reach $f_{NL} \sim O$(a few), providing an interesting target for future CMB missions.  
\end{abstract}

\maketitle

\section{Introduction}
The latest Cosmic Microwave Background (CMB) data \cite{Akrami:2018odb} has confirmed that inflationary perturbations are Gaussian, adiabatic and almost scale-invariant to a high level of accuracy. An elegant explanation for this observed simplicity is that inflation is driven by a single scalar field with small self-interactions. \\

However, inflationary models derived from e.g. supergravity or string theory 
%However, ultraviolet (UV) embeddings of inflation in e.g. supergravity or string theory 
typically contain multiple scalar fields besides the inflaton, often with non-standard kinetic terms \cite{Baumann:2014nda}. Therefore, it is important to understand which properties of multi-field models of inflation lead to predictions compatible with the observational data. At the same time, it would be very desirable to have a simple framework for multi-field inflation in highly curved field spaces and trajectories (see this list for recent developments \cite{Easson:2007dh, Yang:2012bs, Achucarro:2012yr, Brown:2017osf, Achucarro:2017ing, Christodoulidis:2018qdw, Garcia-Saenz:2018ifx,Achucarro:2018vey, Bjorkmo:2019aev, Fumagalli:2019noh, Bjorkmo:2019fls, Christodoulidis:2019mkj, Christodoulidis:2019jsx, Aragam:2019khr}.)\\

The observational data provide us with a glimpse of the Lagrangian of inflationary \textit{perturbations}. This suggests we should classify inflationary models by the behavior of the couplings between perturbations \cite{Cheung:2007st, Chen:2009zp, Baumann:2011nk, Noumi:2012vr, Gwyn:2012mw, Lee:2016vti, Finelli:2018upr, Arkani-Hamed:2018kmz}. An important question is how to relate the simplicity of the perturbations to properties of the full action. The time-dependent background of inflation and the coupling to gravity turn this question into a non-trivial task. For example, to drive slow-roll single-field inflation a non-zero gradient of the potential is needed to compensate for the Hubble friction. The potential therefore necessarily breaks the shift symmetry. Nevertheless, the action for curvature perturbations is shift symmetric. For multi-field scenarios the situation becomes even more complicated.

In its simplest realization, so-called ``canonical single-field slow roll inflation",  the inflaton's evolution follows the gradient flow of the potential, $\dot \phi \sim \nabla_\phi V$ and the inflaton acts as a clock. The power and the beauty of this approach is that knowledge of a single function --the inflaton potential $V(\phi)$-- is enough to predict all the inflationary observables. And viceversa:  the inflaton potential can be constrained by the cosmological data and even reconstructed. We would like a framework that extends at least some of the simplicity and predictability of single-field models to multi-field models with strongly curved manifolds and trajectories, provided these curvatures are reasonably constant.\\

In this work we propose a new method to reconstruct simultaneously, in the simplest possible way, a two-field action and inflationary trajectory that match the observations. The idea is to align the inflationary trajectory with a symmetry in the kinetic terms (an isometry in field space) and use an extension of the Hamilton-Jacobi method \cite{Muslimov:1990be, Salopek:1990jq,Lidsey:1991zp,Copeland:1993jj} to engineer a potential that sustains slow roll {\it multi-field} inflation with the right observables. This sheds new light on non-trivial multi-field structures that are compatible with observational data .\\

The Hamilton-Jacobi formalism for multi-field inflation 
has been previously considered in
\cite{Skenderis:2006jq, Skenderis:2006rr,Byrnes:2009qy,Saffin:2012et,Garriga:2014fda, Garriga:2015tea}
 %SkenderiaTownsend,SkenderisTownsend,ByrnesTasinato,GarrigaSkenderisUrakawa, GarrigaUrakawaVernizzi,Saffin 
 and more recently, in \cite{Achucarro:2018ngj, Achucarro:2019pux}.
% the context of fake supersymmetry \cite{SkenderiaTownsend,SkenderisTownsend} and of holographic inflation \cite{McFaddenSkenderis, GarrigaSkenderisUrakawa, GarrigaUrakawaVernizzi} 
%\cite{GarrigaSkenderisUrakawa, ByrnesTasinato} in analytic estimates of non-gaussianity. Non-canonical kinetic terms were considered in \cite{GarrigaUrakawaVernizzi, Saffin} and more recently, in \cite{ShiftSymm, HolInfl}} \\
The idea is to consider inflationary models that attract to the \textit{Hubble gradient flow} 
$\dot\phi^a \sim - \nabla^a H $, 
%$\dot\vec\phi \sim - \vec\nabla H $, 
which we align with a symmetry in the kinetic terms, i.e. an isometry or `angular' direction $\theta$ in field space. We dub this `Orbital Inflation', since inflaton happens at constant `radius' $\rho = \rho_0$. The non-zero, constant radius ensures the perturbations tangential and orthogonal to the trajectory are coupled and the effective masses and couplings of the perturbations can be made constant, up to slow roll corrections.\\

In single field inflation a single function $V(\varphi)$ determines the behavior of both the background and perturbations. Similarly, in this class of models, once we know the geometry of field space, a single function $H(\phi^a)$ determines the behavior of both the background and perturbations. For instance, the mass of the isocurvature perturbations is fully determined by the hessian of $H$ \cite{Achucarro:2018ngj} \\

In \autoref{sec:kinematicalanalysis} we employ a multi-field generalization of the Hamilton-Jacobi formalism to engineer a class of potentials that sustains slow roll inflation with the right observables, through the simple relation $V=3H^2-2(\nabla^a H)(\nabla_a H)$. These potentials admit \textit{exact} inflationary solutions that follow the Hubble gradient flow and can be studied analytically much like in the single-field case.
%To drive slow-roll inflation the Hubble parameter has an approximate shift symmetry in the isometry direction. Furthermore, the Hessian of the Hubble parameter determines the mass of isocurvature perturbations. 
%Using appropriate angular and radial coordinates, the Hubble parameter (and \textit{not} the potential) takes a particularly simple form for two-field models of inflation, assuming a constant radius of curvature and constant entropy mass.
%As function of the angular and radial coordinates, 
This tells us that the Hubble parameter - and \textit{not} the potential - takes a particularly simple form to support Orbital Inflation, namely $\partial_\rho H =0$ and $\epsilon \sim -\left(\partial_\theta \ln H\right)^2$, both evaluated at $\rho=\rho_0$.  
We can then choose $H(\theta, \rho)$ so that the mass of the isocurvature perturbation (orthogonal to the trajectory)  becomes constant (in units of H) up to a slow roll correction and can be dialled to any desired value between $0$ and $3H/2$. This is, to our knowledge, the first exact realization of quasi-single field inflation \cite{Chen:2009zp, Chen:2009we}, and we exploit it to  a strong turning regime not previously explored. In \cite{Welling:2019qsf}, on the other hand, we test the analytical predictions in the quasi-single field regime.\\

If we consider a product separable Hubble parameter $H(\theta,\rho)=W(\theta)X(\rho)$, with $\partial_\rho X(\rho_0) = 0$ and $\partial_{\rho\rho} X(\rho_0) = \text{constant}$, the mass of the radial isocurvature perturbations 
%becomes constant in units of $H$ up to a slow-roll correction. Moreover, the entropy mass 
is determined by the value of $\partial_{\rho\rho} X(\rho_0)$. This means that the Hubble parameter carries properties that are normally assigned to the potential.  
In particular, if the Hubble parameter has a shift symmetry in the radial field, the isocurvature perturbations become exactly massless. On the other hand, the potential is not shift symmetric in that case.\\

The first order Hubble gradient flow equations allow us to exactly solve for the background dynamics, and therefore also for the superhorizon evolution of inflationary perturbations, if we combine it with the $\delta N$ formalism \cite{Starobinsky:1986fxa, Salopek:1990jq, Sasaki:1995aw, Sasaki:1998ug, Lee:2005bb, Abolhasani:2019cqw}. We work out the details in \autoref{sec:simpleanalyticalestimate}. In \autoref{sec:phenomenology} we use our reconstructed potential to study the phenomenology of two-field inflation with a small entropy mass and a small radius of curvature (a large turn rate), a regime typically not considered in quasi-single-field inflation. We show how single-field like predictions are obtained in the limit, although the single field consistency relations \cite{Maldacena:2002vr, Creminelli:2004yq} are violated. Moreover, we find that the entropy mass dictates how the inflationary predictions fan out in the ($n_s$, $r$) plane.  %Finally, we briefly comment on the possibility of testing the predictions of QSF inflation numerically.   

\section{Inflation along an isometry}\label{sec:kinematicalanalysis}
We focus on two-field inflationary models of the form
\begin{equation} 
 S = \frac{1}{2} \int d^4 x \sqrt{-g}\left[M_p^2 R - G_{ab}\partial_\mu \phi^a \partial^\mu \phi^b - 2V(\phi^a) \right]\ .
\end{equation}  
Here $G_{ab} (\phi^c)$ is the field metric characterizing the kinetic terms. Moreover, $R$ is the Ricci scalar of spacetime and $V(\phi^a)$ the potential energy density of the scalar fields. In Appendix \autoref{app:kinematicalanalysis} we briefly recap the main elements of the kinematical two-field analysis. From now on we will set the (reduced) Planck mass to unity, $M_p^{-2} = 8\pi G = 1$.\\

We are interested in field spaces with an isometry. Without loss of generality we can choose an 'angular' field space coordinate $\theta$ along the isometry,  and a 'radial' coordinate $\rho$ orthogonal to it, the field space metric takes the simple form $\rm{diag}(1, f(\rho))$. The action becomes
\begin{equation}
 S = \frac{1}{2}\int d^4 x \sqrt{-g}  \left[  R -  (\partial_\mu \rho)^2
 -  f(\rho) (\partial_\mu\theta)^2  - 2V(\rho, \theta) \right]
 \end{equation}
 where the   potential $V$ is determined by the Hubble parameter $H (\rho,\theta)$ as
\begin{equation}
V =   3H^2 - (\partial_\rho H)^2 - \frac{(\partial_\theta H)^2}{f(\rho)}      \\ .
\label{eqn:HJpotential}
\end{equation}

The equations of motion for this system are first order, and classical trajectories follow the Hubble gradient flow.
%\subsection{Reconstruction Orbital Inflation}\label{subsec:reconstructionpotential}
%We now construct inflationary potentials with constant radius of curvature, using the Hamilton-Jacobi equation \autoref{eqn:HJtwofield}. 
A constant radius of curvature can be achieved by considering an inflationary trajectory that proceeds along an isometry direction of the field metric that is \textit{not} a geodesic. This is the key characteristic of our reconstruction. We name this class of models `Orbital Inflation'. \\ %And in this sense it is an attempt of realizing spontaneously symmetry probing \cite{Nicolis:2011pv} in inflation. 
%The existence of an isometry implies that we are free to choose field coordinates $(\theta, \rho)$, such that the field metric $G_{ab}$ does not depend on $\theta$. Moreover, we have also the freedom to put $G_{\theta\rho}$ to zero\footnote{If $G_{\theta\rho}\neq 0$, define $\tilde\theta = \theta + \int d\rho\ \frac{G_{\theta\rho}(\rho)}{G_{\theta\theta}(\rho)}$ such that $\tilde G_{\tilde\theta\rho} = 0$.}. Furthermore, we denote $f(\rho) = G_{\theta\theta}$. In other words, we choose coordinates such that the scalar action is
%\begin{equation}
 %S_\phi =-\frac{1}{2} \int d^4 x \sqrt{-g}\left[f(\rho) \partial\theta^2 + \partial \rho^2 + 2V(\phi^a) \right]\ .
%\end{equation}  

%We will derive the local form of the potential for \textit{two} fields. %In this case it turns out to be useful to work with the proper field distance $\varphi$ because it can easily be expressed in terms of $\theta$ on the desired solution. The two equations we should solve simultaneously are the Hamilton-Jacobi equation and constraint equation, respectively:
% \begin{subequations}
% \begin{empheq}{align}
% 3H^2 M_p^2 &= V + 2 M_p^4 H_\varphi^2 \\
% -2M_p^2 \Omega H_\varphi &= N^a V_a.
% \end{empheq}
% \label{eqn:HJeqns}
% \end{subequations}
We would like to reconstruct the potential that admits solutions of the form \begin{equation}
\dot{\rho} = 0 , \quad \dot\theta < 0.                                                                
\end{equation}
The sign of $\dot\theta$ is our choice of convention, also we take $\theta > 0$ on the inflationary trajectory.
%This means we can replace the proper field distance $\varphi \rightarrow \sqrt{f}\theta$ and $H_T  \rightarrow \frac{1}{\sqrt{f}}H_\theta$. Moreover, 
The relevant kinematical and geometrical inflationary quantities\footnote{The precise definitions of these kinematical and geometrical variables can be found in Appendix \ref{app:kinematicalanalysis}.} of Orbital Inflation simplify to 
\begin{subequations}
\begin{empheq}{align}
%&T^a = \frac{1}{\sqrt{f}}(-1,\ 0), \quad N^a  = (0,1), \quad  
&\frac{1}{\kappa} = \frac{f_\rho}{2 f},
%&\dot\s  = -2M_p^2 \frac{H_\theta}{\sqrt{f}},\\
\quad \dot\theta  = -2\frac{H_\theta}{f}, \label{eqn:Hgradientflow}\\
& \epsilon  = \frac{2 H_\theta^2}{f H^2} , \quad \mathbb{R}=\frac{2}{\kappa^2}-\frac{f_{\rho\rho}}{f}, \\
& \mu^2  = 6 H H_{\rho\rho} -4  H_{\rho\rho}^2-\frac{4 H_\theta H_{\theta\rho\rho} }{f}\ .
 \label{eqn:entropymasssol}
\end{empheq}
\label{eqn:exactsolutionbg}
\end{subequations}
We use the shorthand notation $f=f(\rho)$, $H=H(\theta,\rho)$ and $H_\rho = \partial_\rho H$ etcetera.
%Here $T^a$ and $N^a$ are the tangent and normal vector to the inflationary trajectory, respectively. Furthermore, $\kappa$ equals the field radius of curvature, 
Here, $\kappa$ is the turning radius of the trajectory, $\epsilon \equiv  - \frac{\dot H}{H^2}$ is the first slow-roll parameter and $\mathbb{R}$ denotes the Ricci curvature of field space. Finally, $\mu^2$ is the entropy mass \textit{on} the inflationary trajectory $\rho=\rho_0$: the effective mass of the radial isocurvature perturbations. \\

In order to reconstruct a potential that admits an inflationary solution along the angular direction, we need $H_\rho(\rho_0)=0$ for some $\rho_0$, in accordance with the Hubble flow \eqref{eqn:Hgradientflow}. Moreover, for it to be a stable trajectory with respect to perturbations in $\rho$, we take $0 < M_p H_{\rho\rho}(\rho_0)  \lesssim 3H/4$, leading to entropy masses $0 < \mu  \lesssim 3H/2$, as long as the last term in \eqref{eqn:entropymasssol} is slow-roll suppressed (see \cite{Achucarro:2018ngj} for a detailed discussion of the mass bounds in the multi-field case). Other than that we can choose the Hubble parameter as we like and the corresponding potential is given by \eqref{eqn:HJpotential}. Notably, if $H$ (and not $V$) acquires a shift symmetry in $\rho$ , the entropy perturbations are exactly massless. We studied this neutrally stable case elsewhere \cite{Achucarro:2019pux}, and we showed that the predictions for isocurvature and non-gaussianity are strongly suppressed.  \\

Note that we can straightforwardly generalize Orbital Inflation by correcting the Hamilton-Jacobi potential
\begin{equation}
V(\theta, \rho) = V_\text{HJ}(\rho,\theta) + \Delta V(\rho,\theta).
\end{equation}
In order to preserve the background trajectory, i.e. the amount of Hubble friction and the driving force along the trajectory 
%given by \autoref{eqn:fieldandfriedmannequations}, 
the correction term is constrained by %§§,.....
\begin{equation}
\Delta V(\rho_0, \theta) = 0, \quad  \partial_a \Delta V(\rho_0, \theta) =0,
\end{equation}
for all $\theta$,
but other than that we are free to modify the potential. In particular this means Orbital Inflation can accommodate any value of the entropy mass. \\ %\ac{I am not sure we still have an attractor, I got confused about this, but we don't need to mention it} 

%With a simple modification we can generalize this to any value of the entropy mass.\\ 

%\textbf{General mass - } 
%We can go beyond the Hubble flow by adding a contribution $F(\phi^b) N_a$ to the right hand side of \autoref{eqn:HJdynamicstwofield}. The resulting $H_a$ still constitutes a solution to the second Friedmann equation, because the normal contribution is projected out. In our reconstruction we simply choose $F = H_\rho$ such that the condition $\dot\rho=0$ remains valid\footnote{Moreover, the condition $H_\rho(\rho_0) = F_{\rho\rho}(\rho_0) =0$ ensures compatibility with the normal projection of the field equations \autoref{eqn:multifieldequations}.}. The Hamilton-Jacobi equation \autoref{eqn:HJtwofield} receives a correction and becomes
%\begin{equation}
%V = 3H^2M_p^2 - 2 M_p^4 \frac{H_\theta^2}{f},
%\label{eqn:potentialwithanymass}
%\end{equation}
%i.e. the $H_\rho^2$ term disappears. We find that the second contribution to the entropy mass \autoref{eqn:entropymasssol} vanishes. This allows us to reconstruct inflationary models that admit Orbital Inflation with any entropy mass.\\

Let us now consider Orbital Inflation with a constant entropy mass in units of $H$. As can been seen from \autoref{eqn:entropymasssol}, 
%the simplest way 
a simple ansatz to make the entropy mass constant (up to a slow-roll correction) is to take a Hubble parameter of the product separable form 
\begin{equation}
H(\rho, \theta) = W(\theta)X(\rho),\label{eqn:productseparablehubble}
\end{equation}
with $X(\rho_0)=1$, $X_\rho(\rho_0)=0$ and $X_{\rho\rho}(\rho_0)=\lambda M_p^{-2}$. 
Using \autoref{eqn:exactsolutionbg}, the entropy mass is given by $\mu^2/H^2 = 6\lambda -4 \lambda^2-2\epsilon\lambda$.
% \begin{equation}
%  H = W(\theta)\left(1+\lambda\frac{(\rho-\rho_0)^2}{2M_p^2}+\alpha\frac{(\rho-\rho_0)^3}{6M_p^3}+\ldots \right)
%  \label{eqn:reconstructedhubble}
% \end{equation}
A product separable Hubble parameter yields the following subclass of potentials that admit Orbital Inflation
%\begin{align}
 %\frac{V(\rho,\theta)}{M_p^4} = & 3X^2(\rho)\left[\frac{W^2(\theta)}{M_p^2}-\frac{2 W_\theta^2(\theta)}{3f(\rho)}\right] \nonumber \\
% & -2 X_\rho^2(\rho) W^2(\theta)\ .
%\label{eqn:reconstructedpotential}
%\end{align}
\begin{equation}
 \frac{V(\rho,\theta)}{M_p^4} =  3X^2(\rho)\left[\frac{W^2(\theta)}{M_p^2}-\frac{2 W_\theta^2(\theta)}{3f(\rho)}\right] -2 X_\rho^2(\rho) W^2(\theta)\ .
\label{eqn:reconstructedpotential}
\end{equation}

% \begin{widetext}
% \begin{equation}
%  V(\rho,\theta) = 3M_p^2\left(1+\frac{1}{2}\lambda\frac{(\rho-\rho_0)^2}{M_p^2}+\frac{1}{6}\alpha\frac{(\rho-\rho_0)^3}{M_p^3}+\ldots \right)^2\left(W^2(\theta)-\frac{2 M_p^2 W_\theta^2(\theta)}{3f(\rho)}\right)\ .
%  \label{eqn:reconstructedpotential}
% \end{equation}
% \end{widetext}
In addition to the entropy mass, higher order couplings such as $V_{\rho\rho\rho}$ (that is important for non-gaussianity) are determined by the value of $\alpha \equiv X_{\rho\rho\rho}(\rho_0)M_p^3$. We describe the phenomenology of this set-up in the regime of small entropy mass in detail in the next section. \\

Another interesting application of Orbital Inflation is that we can numerically study any realization of quasi-single-field inflation by choosing the parameters in $H(\rho,\theta)$ accordingly. For this we wish to refer the reader to \cite{Welling:2019qsf}.

\section{Phenomenology of Orbital Inflation}\label{sec:phenomenology}
As an example we work out the phenomenology\footnote{All computational details of this analysis can be found in \autoref{sec:simpleanalyticalestimate}. The slightly more general observable predictions are listed in \autoref{eqn:predictionobservables}.} for a particular model of Orbital Inflation in which the Hubble parameter is product separable
\begin{equation}
H(\theta, \rho) = {\cal A} ~ \theta \left(1+ \frac{\lambda}{2} (\rho-\rho_0)^2 + \frac{\alpha}{6}(\rho-\rho_0)^3 + \ldots \right)
\end{equation}
and the field metric is hyperbolic.
%\begin{equation} 
%2K = f(\rho)(\partial\theta)^2+(\partial\rho)^2 \quad \text{with} \quad f(\rho) = e^{2\rho/R_0}\ . 
%\end{equation}
The results in this section are therefore obtained using the following potential and kinetic term
\begin{subequations}
\begin{empheq}{align}
 &V(\theta, \rho) = 3  {\cal A}^2  \left(\theta^2-\frac{2}{3 f(\rho)}\right) \nonumber \\
 & \qquad \qquad \quad  \times \left(1+\frac{\lambda}{2}(\rho-\rho_0)^2+ \frac{\alpha}{6}(\rho-\rho_0)^3 \right)^2 \nonumber \\
 &\qquad  \qquad  - 2 {\cal A}^2 \theta^2 \left(\lambda (\rho-\rho_0)+ \frac{\alpha}{2}(\rho-\rho_0)^2  \right)^2,  \\ 
 & 2K = f(\rho)(\partial\theta)^2+(\partial\rho)^2 \quad \text{with} \quad f(\rho) = e^{2\rho/R_0}\ . 
\end{empheq}
 \label{eqn:numericalpotentialorbitalinflation}
\end{subequations}
The hyperbolic field metric has Ricci curvature $\mathbb{R} = -\frac{2}{R_0^2}$. %As explained in \autoref{subsec:reconstructionpotential},
This model admits a slow-roll  inflationary trajectory with constant turning radius $\kappa^2 = R_0^2$, approximately constant entropy mass $\mu^2/H^2 = 6\lambda - 4\lambda^2 - 2\epsilon \lambda $ and slow-roll parameter $\epsilon = 1/(2\Delta N + 1)$. Note that setting $R_0 \to \infty$ and freezing $\rho = \rho_0$ formally recovers the single-field limit. \\

\begin{figure}[h]
\includegraphics[width=0.45\textwidth]{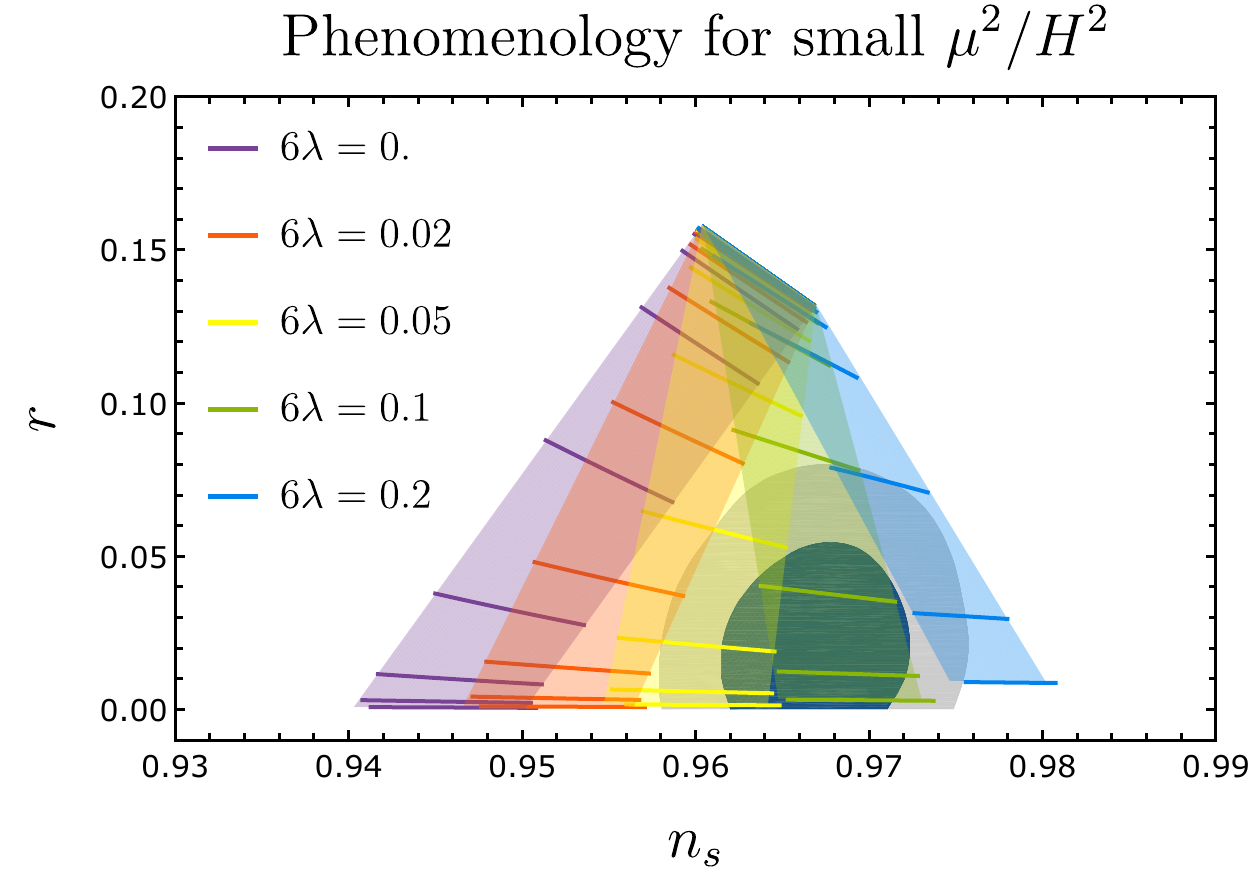}
\centering
\caption{This figure shows the predictions of $(n_s, r)$ at the end of inflation for the model given in \autoref{eqn:numericalpotentialorbitalinflation} using the numerical code \cite{Mulryne:2016mzv, Ronayne:2017qzn}. The entropy mass takes five different values, as indicated in the legend, with $\mu^2/H^2 \approx 6\lambda$. The solid lines correspond to $R_0^2 \in \{1, 4, 4^2, \ldots, 4^8 \}$ from bottom to top, and we let $\Delta N\in \left[50,60\right]$. We plot our analytical results on top (coloured shaded regions) using \autoref{eqn:analyticalnsreps}, where we vary $\kappa^2$ (i.e. $R_0^2$) between $1$ and $10^5$. Furthermore, we plotted the $1\sigma$ and $2\sigma$ confidence contours from \textit{Planck} \cite{Akrami:2018odb} on the background.}
% analytical (coloured shaded regions) and numerical (coloured solid lines) predictions of $(n_s, r)$ for Orbital Inflation. The entropy mass takes five different values, as indicated in the legend, with $\mu^2/H^2 = \lambda\left(1-\frac{1}{3}\epsilon\right) \approx \lambda$. Moreover, to get the analytical contours we use \autoref{eqn:analyticalnsreps}, where we vary $\kappa^2/M_p^2$ between $1$ and $10^5$, and let $\Delta N\in \left[50,60\right]$. The numerical results \cite{Mulryne:2016mzv, Ronayne:2017qzn} are obtained using \autoref{eqn:numericalpotentialorbitalinflation}. The solid lines correspond to $R_0^2/M_p^2 \in \{1, 4, 4^2, \ldots, 4^8 \}$ from bottom to top.  We compare the predictions with the $1\sigma$ and $2\sigma$ confidence contours from \textit{Planck} \cite{Ade:2015lrj}, plotted on the background.}
\label{fig:nsrplotformuandkappa}
\end{figure}

We explore the parameter space with a small but constant entropy mass $0<\mu^2/H^2\ll 9/4$  and a small radius of curvature (but $\kappa^2 \gg 8\epsilon$). This regime is typically not considered in quasi-single field inflation \cite{Chen:2009zp, Chen:2009we}, but is interesting as it shows single-field-like behavior.  
We perform a numerical analysis using the Python code developed by \cite{Mulryne:2016mzv, Ronayne:2017qzn} (see also \cite{Dias:2016rjq, Seery:2016lko}) and plot the predictions for $n_s$ and $r$ at the end of inflation in \autoref{fig:nsrplotformuandkappa}.

We compare our results with an analytical estimate of $n_s$ and $r$ by combining the Hubble gradient flow approximation with the $\delta N$ formalism, please see \autoref{sec:simpleanalyticalestimate} for more details. The power spectrum is given by
\begin{equation}
 P_\R = \frac{H^2}{8\pi^2 \epsilon}\left(1+\dd\right),
 \label{eqn:powerspectrumorbitalinflation}
\end{equation}
where $\dd$ denotes the transfer of power from the isocurvature perturbations to the curvature perturbations and is given by
\begin{equation}
 \dd =  \frac{2\epsilon}{\lambda^2 \kappa^2}\left(1-e^{-2\lambda\Delta N}  \right)^2\ .
 \label{eqn:constantPS}
\end{equation}
%\begin{equation}
 %P_\R \approx \frac{H^2}{8\pi^2 \epsilon}\left(1+2\epsilon \left(\frac{6 H^2}{\mu^2 \kappa}\right)^2\left(1-e^{-\frac{\mu^2}{3H^2}\Delta N}  \right)^2\right)
 %\label{eqn:powerspectrumorbitalinflation}
%\end{equation}
%Here all variables are understood to be evaluated at horizon crossing, and 
Here $\Delta N$ counts the number of efolds between horizon crossing and the end of inflation. %In the limit that $\mu^2/H^2\rightarrow 0$, the second term in \autoref{eqn:powerspectrumorbitalinflation} becomes proportional to $\Delta N^2/\kappa^2$.  
The analytical results 
\begin{equation}
 n_s = 1-\frac{\partial \ln P_\R}{\partial \Delta N}, \quad 
 %r = \frac{2H^2}{\pi^2M_p^2}\frac{1}{P_\R}
 r=\frac{16\epsilon}{1+\dd}, \quad \epsilon = \frac{1}{2\Delta N + 1}
 \label{eqn:analyticalnsreps}\ 
\end{equation}
correspond to the shaded coloured contours in \autoref{fig:nsrplotformuandkappa}.
Notice the excellent agreement with the full numerical computation. In \autoref{fig:nsrplotformuandkappa} we vary $\kappa^2$ between $1$ and $10^5$. In this regime perturbation theory is under control.\\

We find that the observable predictions are already significantly modified with respect to the single-field ones $(\kappa \to \infty, \ \frac{\mu}{H} \to \infty)$ for $\kappa^2 \lesssim 10^3$ (for $\mu^2/H^2=0$) or $\kappa^2 \lesssim 10^2$ (for $\mu^2/H^2= 0.2$). Interestingly, the entropy mass $\mu^2/H^2$ dictates how the inflationary predictions fan out in the $(n_s,r)$ plane. 
%downwards and to the \textit{left} or \textit{right}. \yw{what do we want to say about this? why is this surprising, except for the fact that we haven't often encountered this ourselves?} 
In particular, the various entropy masses predict a different change in $n_s$. It would be very interesting to see if this effect may allow us to distinguish between the various entropy masses. This requires a complementary analysis of the bispectrum, to which we turn now. \\

We are particularly interested in the highly curved regime in which the power spectrum is dominantly sourced by the isocurvature perturbations, $\dd \gg 1$.
In this part of parameter space the power spectrum $P_\s$ of isocurvature  perturbations is suppressed, i.e. $P_\s/P_\R= e^{-2\lambda\Delta N}/(1+\dd) \ll 1 $, where $\s$ is defined here as
$\s \equiv \frac{\delta \rho}{\sqrt{2\epsilon}}$
 (see \autoref{sec:simpleanalyticalestimate}).
Second, because the superhorizon evolution of $\R$ gives the dominant contribution to its final amplitude, we expect that the bispectrum will be of the local type. %\yw{, which we indeed confirm numerically \cite{Mulryne:2016mzv, Ronayne:2017qzn}}. 
In \autoref{sec:simpleanalyticalestimate} we estimate its amplitude. For the hyperbolic field space our analytical prediction becomes
\begin{equation}
f_\text{NL} \approx -\frac{5}{12} \left(\alpha \kappa +6\lambda \right)+\frac{5}{3\kappa}\sqrt{\frac{2\epsilon}{\dd}} \ .
\label{eqn:fnl}
\end{equation}
This estimate only applies for finite $\kappa$ for which $\dd \gg 1$. Its amplitude can take $\O(1)$ values if $\alpha \sim 1 $. If both $\lambda$ and $\alpha$ are of order $O(\epsilon)$, $f_\text{NL}$ becomes slow-roll suppressed, like in single-field inflation. 
We can distinguish Orbital Inflation from single-field inflation though, because the single-field consistency relations \cite{Maldacena:2002vr, Creminelli:2004yq} $f_\text{NL} = \frac{5}{12}(1-n_s)$  and $r=-8n_t = 16\epsilon $ are violated.  \\

Finally, from \autoref{eqn:fnl} we can also compute the spectral tilt of the bispectrum
\begin{equation}
n_{f_\text{NL}} = \frac{-20\epsilon}{3 \kappa^2  f_\text{NL}}\frac{e^{-2\lambda \Delta N}}{\dd}\ .
\end{equation}
Notice that the second slow-roll parameter $\eta$ appears in none of the observables.
Incidentally, if the field metric is known, we can in principle fix all five parameters $\{\lambda, \kappa, \epsilon, \alpha, \Delta N \}$ from observations.

%Finally, we estimated the relative amplitude of isocurvature perturbations compared to the amplitude of curvature perturbations at the end of inflation. How they are related to the late-time non-adiabatic perturbations depends on the mechanism of reheating, but we may still want to ensure that they are suppressed at the end of inflation. This leads to an optional additional constraint
%$\kappa^2 \ll  4\Delta N$ \'{o}r $\mu^2/H^2 \gtrsim 0.1$, which excludes part of the parameter space that we have plotted.\\

%Therefore, it would be interesting to assess the the amplitude and scale-dependence of the bispectrum in the squeezed configuration as a function of $\mu$ and $\kappa$. This we leave to future work. \yw{say something more substantial?}

\section{Final comments}
%Because of the time-dependent background of inflation and the need of a driving force, the inflationary potential does not necessarily reflect the simplicity of the perturbations. For example a shift symmetric potential cannot drive slow-roll inflation. Nevertheless, the action for curvature perturbations is shift symmetric. For inflation with more fields the situation becomes even more complex. When we allow for multiple scalar degrees of freedom that interact, the properties of the scalar potential becomes quite intractable. 

In this work we have reconstructed exact models of two-field inflation that exploit an isometry in field space to get  the simplest possible quadratic action of perturbations (with constant coefficients). We find that the simplicity of the perturbations -- especially those orthogonal to the trajectory -- is manifest in the Hubble parameter $H(\phi^a)$, and not in the inflationary potential $V(\phi^a)$. This is tied to the fact that the inflationary trajectories we consider are generically \textit{not} aligned with the potential gradient flow.  An important question is to understand the (approximate) symmetries of the perturbations as a consequence of an (approximate) symmetry of the full UV theory. In the case of the hyperbolic field space we can connect the behavior of perturbations to a scaling symmetry \cite{Achucarro:2018def}. \\
%This improves our understanding of building blocks of multi-field inflation that respect the simple behavior of inflationary perturbations that we observe.\\

%The icing on the cake is that we have at hand 
We have identified a family of \emph{exactly solvable} two-field models that provide textbook case studies for quasi-single-field inflation. Their dynamics is exact, and the mass and interactions of perturbations can be 
%tuned in any way one likes
dialled to any desired values, within some ranges. In \cite{Welling:2019qsf} we exploit this fact to test the predictions of quasi-single-field inflation. \\

%Applying the delta N formalism... can compute power spectrum and bispectrum...   see Appendix XX
%Summarizing, 
%We reproduce here the predictions for the observables 
In this work we focus on the regime of small entropy mass, and analytically solve for the observables. Details can be found in appendix 
\ref{sec:coupledperturbations}, we reproduce here the results for completeness.
In the limit $\dd \gg 1$ we find the following predictions for the tensor-to-scalar ratio, the spectral index, the tensor tilt, the amplitude of the reduced bispectrum, and its tilt, respectively:  
 \begin{subequations}
\begin{align}
  &r = \frac{16\epsilon}{1+\dd},\\
%  &n_s = 1 - 2\epsilon - 4\lambda e^{-2\lambda  \Delta N}\left(1-%e^{-2\lambda  \Delta N}\right)^{-1},\\
& n_s = 1 -2 \epsilon - \frac{1-4\kappa \lambda N_\rho}{N_\rho}, \\
  &n_t = -2\epsilon,\\
  &f_\text{NL} = -\frac{5}{12} \left(\alpha \kappa +\lambda \frac{10-\mathbb{R} \kappa^2}{2}\right)+\frac{5}{N_\rho}\frac{2-\mathbb{R} \kappa^2}{12\kappa},\\
 & n_{f_\text{NL}} \equiv \frac{d f_\text{NL}/ d\Delta N}{f_\text{NL}} = \frac{2-\mathbb{R} \kappa^2}{\kappa^2}\frac{1-\kappa \lambda N_\rho}{N_{\rho\rho}} \ ,
\end{align}
\label{eqn:predictionobservables}
\end{subequations}
with $\dd = 2\epsilon N_\rho^2$, and $N_\rho$ and $N_{\rho\rho}$ are given by
\begin{align}
%&\frac{N_\theta}{\sqrt{f_0}} = \sqrt{\frac{1}{2\epsilon}}, \\
&N_\rho = \frac{1}{\kappa\lambda }\left(1-e^{-2\lambda \Delta N} \right)\label{eqn:Nrho} , \\
& N_{\rho\rho} =-N_\rho^2 \left(\frac{\alpha \kappa}{2}+\lambda \frac{10-\mathbb{R} \kappa^2}{4}\right)+N_\rho\frac{2-\mathbb{R} \kappa^2}{2\kappa}.
\end{align} 
% in \autoref{eqn:Nrho} and \autoref{eqn:Nrhorho}. 

%We check these expressions numerically using \textit{PyTransport} \cite{Mulryne:2016mzv, Ronayne:2017qzn} (see also \cite{Dias:2016rjq, Seery:2016lko}). 

We showed non-trivial examples of phenomenologically successful models of two-field inflation in the regime where the field manifold and the trajectory can be strongly curved. These models provide an efficient transfer of non-gaussianity from isocurvature self-interactions to the curvature bispectrum. Depending on the size of the isocurvature self-interaction (the parameter $\alpha$) the amplitude of the squeezed bispectrum $f_{NL}$ can range from being slow-roll suppressed to  $\O(1)$, while being consistent with all current observations.
%, providing an interesting target for future CMB space missions such as \cite{PICO1902.10541, Core??}. } \ac{check Core}

%To connect our work to realizing inflation in string theory, an important question is how the desired properties of the Hubble parameter translate to properties of the superpotential in $\mathcal{N}=1$ supergravity with a single chiral superfield. 
%This is a challenging task as potentials reconstructed by Hamilton-Jacobi formalism are of the fake supergravity form, which look similar, but are not equal to the $F$-term potential.  Moreover, it would be interesting to identify the properties of the inflationary potentials that admit solutions approximately equal to the exactly solvable trajectories we constructed in this work. However, the inverse problem of reconstructing the Hubble parameter from the potential is highly non-trivial in two field inflation, as the gradient of the potential is not directly related to the kinetic energy of the scalar fields. 
%Therefore, we leave these interesting questions to future work [CITE future paper of Ana with Sebastian et al].
 
\section{Acknowledgments}
We wish to thank Sebastian Cespedes, Gonzalo Palma, Gianmassimo Tasinato, Dong-Gang Wang and Alexander Westphal for fruitful discussions.
The work of AA is partially supported by the Netherlands' Organization for Fundamental Research in Matter (FOM), by the Basque Government (IT-979-16) and by the Spanish Ministry MINECO (FPA2015-64041-C2-1P). YW is supported by the ERC Consolidator Grant STRINGFLATION under the HORIZON 2020 grant agreement no. 647995.

\bibliographystyle{hieeetr}
\bibliography{bibfile}

\begin{thebibliography}{10}

\bibitem{Achucarro:2019pux}
A.~Ach\'{u}carro, E.~J. Copeland, O.~Iarygina, G.~A. Palma, D.-G. Wang, and
  Y.~Welling, ``{Shift-Symmetric Orbital Inflation: single field or
  multi-field?},'' 2019, 1901.03657.

\bibitem{Achucarro:2016fby}
A.~Ach\'{u}carro, V.~Atal, C.~Germani, and G.~A. Palma, ``{Cumulative effects
  in inflation with ultra-light entropy modes},'' {\em JCAP}, vol.~1702,
  no.~02, p.~013, 2017, 1607.08609.

\bibitem{Chen:2009zp}
X.~Chen and Y.~Wang, ``{Quasi-Single Field Inflation and Non-Gaussianities},''
  {\em JCAP}, vol.~1004, p.~027, 2010, 0911.3380.

\bibitem{Chen:2009we}
X.~Chen and Y.~Wang, ``{Large non-Gaussianities with Intermediate Shapes from
  Quasi-Single Field Inflation},'' {\em Phys. Rev.}, vol.~D81, p.~063511, 2010,
  0909.0496.

\bibitem{Akrami:2018odb}
Y.~Akrami {\em et~al.}, ``{Planck 2018 results. X. Constraints on inflation},''
  2018, 1807.06211.

\bibitem{Baumann:2014nda}
D.~Baumann and L.~McAllister, {\em {Inflation and String Theory}}.
\newblock Cambridge Monographs on Mathematical Physics, Cambridge University
  Press, 2015, 1404.2601.

\bibitem{Easson:2007dh}
D.~A. Easson, R.~Gregory, D.~F. Mota, G.~Tasinato, and I.~Zavala,
  ``{Spinflation},'' {\em JCAP}, vol.~0802, p.~010, 2008, 0709.2666.

\bibitem{Yang:2012bs}
I.-S. Yang, ``{The Strong Multifield Slowroll Condition and Spiral
  Inflation},'' {\em Phys. Rev.}, vol.~D85, p.~123532, 2012, 1202.3388.

\bibitem{Achucarro:2012yr}
A.~Achucarro, V.~Atal, S.~Cespedes, J.-O. Gong, G.~A. Palma, and S.~P. Patil,
  ``{Heavy fields, reduced speeds of sound and decoupling during inflation},''
  {\em Phys. Rev.}, vol.~D86, p.~121301, 2012, 1205.0710.

\bibitem{Brown:2017osf}
A.~R. Brown, ``{Hyperinflation},'' 2017, 1705.03023.

\bibitem{Achucarro:2017ing}
A.~Ach\'{u}carro, R.~Kallosh, A.~Linde, D.-G. Wang, and Y.~Welling,
  ``{Universality of multi-field $\alpha$-attractors},'' {\em JCAP}, vol.~1804,
  no.~04, p.~028, 2018, 1711.09478.

\bibitem{Christodoulidis:2018qdw}
P.~Christodoulidis, D.~Roest, and E.~I. Sfakianakis, ``{Angular inflation in
  multi-field ${\alpha}$-attractors},'' 2018, 1803.09841.

\bibitem{Garcia-Saenz:2018ifx}
S.~Garcia-Saenz, S.~Renaux-Petel, and J.~Ronayne, ``{Primordial fluctuations
  and non-Gaussianities in sidetracked inflation},'' {\em JCAP}, vol.~1807,
  no.~07, p.~057, 2018, 1804.11279.

\bibitem{Achucarro:2018vey}
A.~Achúcarro and G.~A. Palma, ``{The string swampland constraints require
  multi-field inflation},'' {\em JCAP}, vol.~1902, p.~041, 2019, 1807.04390.

\bibitem{Bjorkmo:2019aev}
T.~Bjorkmo and M.~C.~D. Marsh, ``{Hyperinflation generalised: from its
  attractor mechanism to its tension with the `swampland conjectures'},'' 2019,
  1901.08603.

\bibitem{Fumagalli:2019noh}
J.~Fumagalli, S.~Garcia-Saenz, L.~Pinol, S.~Renaux-Petel, and J.~Ronayne,
  ``{Hyper non-Gaussianities in inflation with strongly non-geodesic motion},''
  2019, 1902.03221.

\bibitem{Bjorkmo:2019fls}
T.~Bjorkmo, ``{The rapid-turn inflationary attractor},'' 2019, 1902.10529.

\bibitem{Christodoulidis:2019mkj}
P.~Christodoulidis, D.~Roest, and E.~Sfakianakis, ``{Attractors, Bifurcations
  and Curvature in Multi-field Inflation},'' 2019, 1903.03513.

\bibitem{Christodoulidis:2019jsx}
P.~Christodoulidis, D.~Roest, and E.~I. Sfakianakis, ``{Scaling attractors in
  multi-field inflation},'' 2019, 1903.06116.

\bibitem{Aragam:2019khr}
V.~Aragam, S.~Paban, and R.~Rosati, ``{Multi-field Inflation in High-Slope
  Potentials},'' 2019, 1905.07495.

\bibitem{Cheung:2007st}
C.~Cheung, P.~Creminelli, A.~L. Fitzpatrick, J.~Kaplan, and L.~Senatore, ``{The
  Effective Field Theory of Inflation},'' {\em JHEP}, vol.~03, p.~014, 2008,
  0709.0293.

\bibitem{Baumann:2011nk}
D.~Baumann and D.~Green, ``{Signatures of Supersymmetry from the Early
  Universe},'' {\em Phys. Rev.}, vol.~D85, p.~103520, 2012, 1109.0292.

\bibitem{Noumi:2012vr}
T.~Noumi, M.~Yamaguchi, and D.~Yokoyama, ``{Effective field theory approach to
  quasi-single field inflation and effects of heavy fields},'' {\em JHEP},
  vol.~06, p.~051, 2013, 1211.1624.

\bibitem{Gwyn:2012mw}
R.~Gwyn, G.~A. Palma, M.~Sakellariadou, and S.~Sypsas, ``{Effective field
  theory of weakly coupled inflationary models},'' {\em JCAP}, vol.~1304,
  p.~004, 2013, 1210.3020.

\bibitem{Lee:2016vti}
H.~Lee, D.~Baumann, and G.~L. Pimentel, ``{Non-Gaussianity as a Particle
  Detector},'' {\em JHEP}, vol.~12, p.~040, 2016, 1607.03735.

\bibitem{Finelli:2018upr}
B.~Finelli, G.~Goon, E.~Pajer, and L.~Santoni, ``{The Effective Theory of
  Shift-Symmetric Cosmologies},'' {\em JCAP}, vol.~1805, no.~05, p.~060, 2018,
  1802.01580.

\bibitem{Arkani-Hamed:2018kmz}
N.~Arkani-Hamed, D.~Baumann, H.~Lee, and G.~L. Pimentel, ``{The Cosmological
  Bootstrap: Inflationary Correlators from Symmetries and Singularities},''
  2018, 1811.00024.

\bibitem{Muslimov:1990be}
A.~G. Muslimov, ``{On the Scalar Field Dynamics in a Spatially Flat Friedman
  Universe},'' {\em Class. Quant. Grav.}, vol.~7, pp.~231--237, 1990.

\bibitem{Salopek:1990jq}
D.~S. Salopek and J.~R. Bond, ``{Nonlinear evolution of long wavelength metric
  fluctuations in inflationary models},'' {\em Phys. Rev.}, vol.~D42,
  pp.~3936--3962, 1990.

\bibitem{Lidsey:1991zp}
J.~E. Lidsey, ``{The Scalar field as dynamical variable in inflation},'' {\em
  Phys. Lett.}, vol.~B273, pp.~42--46, 1991.

\bibitem{Copeland:1993jj}
E.~J. Copeland, E.~W. Kolb, A.~R. Liddle, and J.~E. Lidsey, ``{Reconstructing
  the inflation potential, in principle and in practice},'' {\em Phys. Rev.},
  vol.~D48, pp.~2529--2547, 1993, hep-ph/9303288.

\bibitem{Skenderis:2006jq}
K.~Skenderis and P.~K. Townsend, ``{Hidden supersymmetry of domain walls and
  cosmologies},'' {\em Phys. Rev. Lett.}, vol.~96, p.~191301, 2006,
  hep-th/0602260.

\bibitem{Skenderis:2006rr}
K.~Skenderis and P.~K. Townsend, ``{Hamilton-Jacobi method for curved domain
  walls and cosmologies},'' {\em Phys. Rev.}, vol.~D74, p.~125008, 2006,
  hep-th/0609056.

\bibitem{Byrnes:2009qy}
C.~T. Byrnes and G.~Tasinato, ``{Non-Gaussianity beyond slow roll in
  multi-field inflation},'' {\em JCAP}, vol.~0908, p.~016, 2009, 0906.0767.

\bibitem{Saffin:2012et}
P.~M. Saffin, ``{The covariance of multi-field perturbations, pseudo-susy and
  $f_{NL}$},'' {\em JCAP}, vol.~1209, p.~002, 2012, 1203.0397.

\bibitem{Garriga:2014fda}
J.~Garriga, K.~Skenderis, and Y.~Urakawa, ``{Multi-field inflation from
  holography},'' {\em JCAP}, vol.~1501, no.~01, p.~028, 2015, 1410.3290.

\bibitem{Garriga:2015tea}
J.~Garriga, Y.~Urakawa, and F.~Vernizzi, ``{$\delta N$ formalism from
  superpotential and holography},'' {\em JCAP}, vol.~1602, no.~02, p.~036,
  2016, 1509.07339.

\bibitem{Achucarro:2018ngj}
A.~Ach\'{u}carro, S.~Cespedes, A.-C. Davis, and G.~A. Palma, ``{Constraints on
  holographic multi-field inflation},'' 2018, 1809.05341.

\bibitem{Welling:2019qsf}
Y.~Welling, ``{A simple exact model of quasi-single-field inflation},'' {In
  Preparation}.

\bibitem{Starobinsky:1986fxa}
A.~A. Starobinsky, ``{Multicomponent de Sitter (Inflationary) Stages and the
  Generation of Perturbations},'' {\em JETP Lett.}, vol.~42, pp.~152--155,
  1985.
\newblock [Pisma Zh. Eksp. Teor. Fiz.42,124(1985)].

\bibitem{Sasaki:1995aw}
M.~Sasaki and E.~D. Stewart, ``{A General analytic formula for the spectral
  index of the density perturbations produced during inflation},'' {\em Prog.
  Theor. Phys.}, vol.~95, pp.~71--78, 1996, astro-ph/9507001.

\bibitem{Sasaki:1998ug}
M.~Sasaki and T.~Tanaka, ``{Superhorizon scale dynamics of multiscalar
  inflation},'' {\em Prog. Theor. Phys.}, vol.~99, pp.~763--782, 1998,
  gr-qc/9801017.

\bibitem{Lee:2005bb}
H.-C. Lee, M.~Sasaki, E.~D. Stewart, T.~Tanaka, and S.~Yokoyama, ``{A New delta
  N formalism for multi-component inflation},'' {\em JCAP}, vol.~0510, p.~004,
  2005, astro-ph/0506262.

\bibitem{Abolhasani:2019cqw}
A.~A. Abolhasani, H.~Firouzjahi, A.~Naruko, and M.~Sasaki, {\em {Delta N
  Formalism in Cosmological Perturbation Theory}}.
\newblock WSP, 2019.

\bibitem{Maldacena:2002vr}
J.~M. Maldacena, ``{Non-Gaussian features of primordial fluctuations in single
  field inflationary models},'' {\em JHEP}, vol.~05, p.~013, 2003,
  astro-ph/0210603.

\bibitem{Creminelli:2004yq}
P.~Creminelli and M.~Zaldarriaga, ``{Single field consistency relation for the
  3-point function},'' {\em JCAP}, vol.~0410, p.~006, 2004, astro-ph/0407059.

\bibitem{Mulryne:2016mzv}
D.~J. Mulryne and J.~W. Ronayne, ``{PyTransport: A Python package for the
  calculation of inflationary correlation functions},'' 2016, 1609.00381.

\bibitem{Ronayne:2017qzn}
J.~W. Ronayne and D.~J. Mulryne, ``{Numerically evaluating the bispectrum in
  curved field-space— with PyTransport 2.0},'' {\em JCAP}, vol.~1801, no.~01,
  p.~023, 2018, 1708.07130.

\bibitem{Dias:2016rjq}
M.~Dias, J.~Frazer, D.~J. Mulryne, and D.~Seery, ``{Numerical evaluation of the
  bispectrum in multiple field inflation—the transport approach with code},''
  {\em JCAP}, vol.~1612, no.~12, p.~033, 2016, 1609.00379.

\bibitem{Seery:2016lko}
D.~Seery, ``{CppTransport: a platform to automate calculation of inflationary
  correlation functions},'' 2016, 1609.00380.

\bibitem{Achucarro:2018def}
A.~Ach\'{u}carro, G.~A. Palma, D.-G. Wang, and Y.~Welling, ``{On the origin of
  ultra-light fields during inflation and their primordial non-Gaussianity},''
  {In Preparation}.

\bibitem{GrootNibbelink:2000vx}
S.~Groot~Nibbelink and B.~J.~W. van Tent, ``{Density perturbations arising from
  multiple field slow roll inflation},'' 2000, hep-ph/0011325.

\bibitem{GrootNibbelink:2001qt}
S.~Groot~Nibbelink and B.~J.~W. van Tent, ``{Scalar perturbations during
  multiple field slow-roll inflation},'' {\em Class. Quant. Grav.}, vol.~19,
  pp.~613--640, 2002, hep-ph/0107272.

\bibitem{Achucarro:2010da}
A.~Achucarro, J.-O. Gong, S.~Hardeman, G.~A. Palma, and S.~P. Patil,
  ``{Features of heavy physics in the CMB power spectrum},'' {\em JCAP},
  vol.~1101, p.~030, 2011, 1010.3693.

\bibitem{Cespedes:2013rda}
S.~Cespedes and G.~A. Palma, ``{Cosmic inflation in a landscape of
  heavy-fields},'' {\em JCAP}, vol.~1310, p.~051, 2013, 1303.4703.

\bibitem{Lyth:2005fi}
D.~H. Lyth and Y.~Rodriguez, ``{The Inflationary prediction for primordial
  non-Gaussianity},'' {\em Phys. Rev. Lett.}, vol.~95, p.~121302, 2005,
  astro-ph/0504045.

\bibitem{Seery:2005gb}
D.~Seery and J.~E. Lidsey, ``{Primordial non-Gaussianities from multiple-field
  inflation},'' {\em JCAP}, vol.~0509, p.~011, 2005, astro-ph/0506056.

\bibitem{Langlois:2010vx}
D.~Langlois and F.~Vernizzi, ``{A geometrical approach to nonlinear
  perturbations in relativistic cosmology},'' {\em Class. Quant. Grav.},
  vol.~27, p.~124007, 2010, 1003.3270.

\bibitem{Gong:2011uw}
J.-O. Gong and T.~Tanaka, ``{A covariant approach to general field space metric
  in multi-field inflation},'' {\em JCAP}, vol.~1103, p.~015, 2011, 1101.4809.
\newblock [Erratum: JCAP1202,E01(2012)].

\bibitem{Elliston:2012ab}
J.~Elliston, D.~Seery, and R.~Tavakol, ``{The inflationary bispectrum with
  curved field-space},'' {\em JCAP}, vol.~1211, p.~060, 2012, 1208.6011.

\bibitem{Gordon:2000hv}
C.~Gordon, D.~Wands, B.~A. Bassett, and R.~Maartens, ``{Adiabatic and entropy
  perturbations from inflation},'' {\em Phys. Rev.}, vol.~D63, p.~023506, 2001,
  astro-ph/0009131.

\end{thebibliography}
\newpage

\section{Kinematical two-field analysis}\label{app:kinematicalanalysis}
In this appendix we recap the main elements of the kinematical analysis of two-field inflationary models of the form 
\begin{equation}
 S = \frac{1}{2} \int d^4 x \sqrt{-g}\left[M_p^2 R - G_{ab}\partial_\mu \phi^a \partial^\mu \phi^b - 2V(\phi^a) \right]\ .
\end{equation}  
Here $G_{ab}$ is the field metric characterizing the kinetic terms. Moreover, $R$ is the Ricci scalar of spacetime and $V(\phi^a)$ the potential energy density of the scalar fields. We then generalize the Hamilton-Jacobi formalism to two-field inflation, which we employ to derive exact models of Orbital Inflation, featuring a constant radius of curvature. \\

The background dynamics of the scalar fields follows from assuming a homogeneous, isotropic and flat Friedmann Lemaitre Robertson Walker spacetime $ds^2 = -dt^2 + a^2(t)d\v{x}^2$.
The field equations and Friedmann equations are given by \cite{GrootNibbelink:2000vx}
\begin{subequations}
 \begin{empheq}{align}
& D_t^2\phi^a +3H D_t \phi^a + \nabla^a V = 0, \label{eqn:multifieldequations} \\
& 3H^2M_p^2 = \frac{1}{2}G_{ab} \dot\phi^a\dot\phi^b + V(\phi^a), \label{eqn:Friedmannequation}\\
& \dot H M_p^2 = \frac{1}{2}G_{ab} \dot\phi^a\dot\phi^b , \label{eqn:2ndFriedmannequation}
 \end{empheq}
 \label{eqn:fieldandfriedmannequations}
 \end{subequations}
respectively, with $H\equiv \dot{a}/a$ the Hubble parameter. We write $D_t \equiv \dot\phi^a\nabla_a$, with $\nabla_a$ the covariant field derivative with respect to the field metric. Moreover, the latin field indices are raised and lowered with the field metric. The Hubble slow-roll parameter $\epsilon$ is a measure of the kinetic energy of the scalar fields
\begin{equation}
 \epsilon \equiv -\frac{\dot H}{H^2} = \frac{1}{2} \frac{\dot\varphi^2}{H^2 M_p^2},
\end{equation}
with $ \dot\varphi  \equiv  \left(G_{ab} \dot\phi^a\dot\phi^b\right)^{1/2}$ the \textit{proper field velocity}. We wish to emphasize that $\epsilon$ is not a measure of the gradient of the potential, because in general  $\dot\phi^a$ is \textit{not} aligned with $V^a$.\\

The inflationary background trajectory determines a natural basis of unit vectors in field space \cite{GrootNibbelink:2001qt, Achucarro:2010da,Cespedes:2013rda}, reminiscent of the Frenet-Serret equations
\begin{equation}
 T^a = \frac{\dot\phi^a}{\dot\varphi}, \qquad D_t T^a = -\Omega N^a
 \label{eqn:unitbasisvectors}
 \end{equation}
Here $T^a$ is the tangent vector pointing along the inflationary curve, with $ \dot\varphi  \equiv  \left(G_{ab} \dot\phi^a\dot\phi^b\right)^{1/2} $ the proper field velocity.  Furthermore $N^a$ is the normal vector, normalized to unity. This uniquely determines the value of the \textit{turn rate} $\Omega$ up to a sign. Equivalently, we can compute the \textit{radius of curvature} $\kappa$ of any inflationary curve \begin{equation}
 \kappa \equiv  \left(N_a T^b \nabla_b T^a\right)^{-1}\ .
 \label{eqn:radiusofcurvaturedef}
\end{equation}
Notice that the radius of curvature is related to the turn rate as 
$\Omega = -\sqrt{2\epsilon} M_p H \kappa^{-1}$.\\

In two-field inflation the dynamics of linear inflationary perturbations depends, in addition to $\epsilon$ and $\kappa$, on the mass of perturbations, the \textit{entropy mass} $\mu$
\begin{equation}
 \mu^2 \equiv V_{NN}+\epsilon H^2 M_p^2 \mathbb{R} +\frac{6\epsilon M_p^2 H^2}{\kappa^{2}},
 \label{eqn:entropymass}
\end{equation}
where $\mathbb{R}$ is the Ricci scalar of field space and we used the notation $V_{NN}\equiv N^a N^b \nabla_a \nabla_b V$. The definition of the entropy mass follows from a dispersion relation analysis \cite{Achucarro:2012yr} of the coupled system of perturbations (described in \autoref{eqn:quadraticactionperturbationsorbitalinflation}). The time-dependence of the inflationary background induces the geometrical and centrifugal contributions to the entropy mass.

\section{Hamilton-Jacobi formalism}
\subsection{Hamilton-Jacobi for two fields}\label{sec:hamiltonjacobi}
To reconstruct the potential in the neighbourhood of a given background trajectory we use a generalization of the Hamilton-Jacobi formalism to two-field inflation with non-canonical kinetic terms. In this formalism one interprets the Hubble parameter as a function of the field coordinates $H = H(\phi^a)$. This requires the trajectory not to intersect itself. We assume that the field velocity $\dot\phi^a$  is non-zero along the trajectory. %to be monotonically non-decrreasing increasing or decreasing along the inflationary trajectory. %An additional constraint in the two-field scenario is that the gradient of the potential in the orthogonal direction should counterbalance the centrifugal force. \\
%In single field inflation each background solution corresponds to some initial value $H(\phi^a_0)$. In two-field inflation, however, this cannot cover all possible initial data and we need an additional function $F(\phi^a)$ to fully specify the system. Indeed,  
This means we can write $\dot{H} = \dot\phi^a \partial_a H$ and the simplest solution to the second Friedmann equation \autoref{eqn:2ndFriedmannequation} is given by
\begin{equation}
\partial_a H = -\frac{G_{ab}\dot\phi^b}{2M_p^2}\ . %+ F(\phi^b)N_a \ .
 \label{eqn:HJdynamicstwofield}
\end{equation}
%We have the freedom to add to $H_a$ any contribution proportional to the normal vector to the trajectory, because it is projected out when we contract it with $\dot\phi^a$ to get $\dot H$. 
Inflationary trajectories of this kind follow the \textit{Hubble gradient flow}, given by $\dot\phi^a \sim -\nabla^a H$.  
They allow us to rewrite the first Friedmann equation \autoref{eqn:Friedmannequation} as the multi-field Hamilton-Jacobi equation
% \begin{equation}
% V = 3H^2M_p^2 - 2 M_p^4 \left(H^a H_a -F^2\right).
% \label{eqn:HJtwofield}
% \end{equation}
\begin{equation}
V = 3H^2M_p^2 - 2 M_p^4 G^{ab} (\partial_a H) (\partial_b H) .
\label{eqn:HJtwofield}
\end{equation}
The Hamilton-Jacobi equation allows us to reconstruct a potential that admits a given inflationary trajectory as a solution\footnote{As long as there is solution for the Hubble function that is strictly monotonic in all coordinates.}. We immediately see that, generically, the reconstructed potential does not respect the same symmetries as the Hubble parameter. For instance, if the Hubble parameter has a shift symmetry in one of the fields, this may be violated for the potential if the field metric is non-trivial.\\ %(through the last term $H^a H_a = H_a H_b G^{ab}$). \\

%Notice that we need both $H$ and $F$ to fully determine the potential. 
It is straightforward to check that the Hamilton-Jacobi equation \autoref{eqn:HJtwofield} together with the Hubble flow \autoref{eqn:HJdynamicstwofield} is consistent with both the tangent and the normal projection of the field equations \autoref{eqn:multifieldequations}. %Moreover, the Hubble flow ensures 
%we should check that the gradient of the potential in the orthogonal direction counterbalances the centrifugal force., i.e. we have to obey 
%that the normal projection of the field equations is automatically obeyed. 
% This provides the constraint equation 
% \begin{equation}
%  3 H F - 2 M_p^2 H_T F_T  = 0 \ .
%  \label{eqn:constrainttwofieldsimple}
% \end{equation}
% Here we introduce the notation $H_T \equiv T^a H_a$ and likewise for $F_T$. 
% In particular, $F = 0$ solves \autoref{eqn:constrainttwofieldsimple}. In this case all inflationary solutions attract to the \textit{Hubble flow}, given by $\dot{\phi}^a \sim H^a$. \yw{say something about attractors?} However, taking $F=0$ necessarily bounds the entropy mass from above [CITE]. Therefore, if we aim to construct two-field models with a larger entropy mass, we should relax the assumption of Hubble flow. %Since we are interested in a local reconstruction of the potential only, this does not constitute much of a problem.

\subsection{inflating along an isometry}
The existence of an isometry implies that we are free to choose field coordinates $(\theta, \rho)$, such that the field metric $G_{ab}$ does not depend on $\theta$. Moreover, we have also the freedom to put $G_{\theta\rho}$ to zero\footnote{If $G_{\theta\rho}\neq 0$, define $\tilde\theta = \theta + \int d\rho\ \frac{G_{\theta\rho}(\rho)}{G_{\theta\theta}(\rho)}$ such that $\tilde G_{\tilde\theta\rho} = 0$.}. Furthermore, we denote $f(\rho) = G_{\theta\theta}$. In other words, we choose coordinates such that the scalar action is
\begin{equation}
 S_\phi =-\frac{1}{2} \int d^4 x \sqrt{-g}\left[f(\rho) \partial\theta^2 + \partial \rho^2 + 2V(\phi^a) \right]\ .
\end{equation}

\section{Analytical Estimates of the Observables}\label{sec:simpleanalyticalestimate}
In this appendix we provide two simple analytical derivations of the observables of Orbital Inflation. The first approach exploits the Hubble gradient flow to solve for $\rho(N)$ and $\theta(N)$. Consequently we use the $\delta N$ formalism to approximate the power spectrum of curvature perturbations that is generated on superhorizon scales. Moreover, this allows us to estimate the amplitude of primordial non-Gaussianities. The second approach solves the system of coupled perturbations directly, exploiting the fact that the entropy mass and radius of curvature are constant. This has the advantage that we can also estimate the power spectrum of isocurvature perturbations. \\

As in the main text we work with the angular and radial field coordinates suitable to describe Orbital Inflation. The field metric $G_{ab}$ is then diagonal and given by $G_{\rho\rho}=1$ and $G_{\theta\theta}=f(\rho)$. Furthermore, we assume the product separable form for the Hubble parameter $H(\theta,\rho)=X(\rho)W(\theta)$.  \\

For completeness, the following formulas from the main text are useful below:
\begin{subequations}
\begin{empheq}{align}
&T^a = \frac{1}{\sqrt{f}}(-1,\ 0), \quad N^a  = (0,1), \quad  
\frac{1}{\kappa} = \frac{f_\rho}{2 f}, \\
%&\dot\s  = -2M_p^2 \frac{H_\theta}{\sqrt{f}},\\
&\dot\theta  = -2\frac{H_\theta}{f}, \epsilon  = \frac{2 H_\theta^2}{f H^2} , \quad \mathbb{R}=\frac{2}{\kappa^2}-\frac{f_{\rho\rho}}{f}, \\
& \mu^2  = 6 M_p^2 H H_{\rho\rho} -4 M_p^4 H_{\rho\rho}^2-\frac{4M_p^2 H_\theta H_{\theta\rho\rho} }{f}\ .
\end{empheq}
\end{subequations}

\subsection{Hubble gradient flow combined with the $\delta N$ formalism}
We use the Hubble gradient flow \autoref{eqn:HJdynamicstwofield} to solve exactly for the background dynamics of Orbital Inflation. %\yw{change this paragraph} Strictly speaking, in this approach, we are considering the potential \autoref{eqn:HJtwofield} instead of \autoref{eqn:reconstructedpotential}. For the dynamics of perturbations we expect only the values of the couplings to matter, however. 
%Notice from the discussion below \autoref{eqn:entropymasssol} that this forces us to consider entropy masses of $\mu^2/H^2 \leq 9/4$ only. This is no problem, since in \autoref{sec:phenomenology} we are mainly interested in $\mu^2/H^2\ll 9/4$. 
Since the Hamilton-Jacobi function is product separable $H(\theta,\rho) = W(\theta)X(\rho)$, the formal solution to the Hubble gradient flow \autoref{eqn:HJdynamicstwofield} for this class of Orbital Inflation is given by
\begin{align}
&\int_{\theta_\text{in}}^{\theta_\text{end}} \frac{d\theta}{\partial_\theta(\ln W(\theta))} = -\int_{0}^{\Delta N}  \frac{2}{f(\rho)} dN\ ,\label{eqn:thetasolutionHJ}\\
&\int_{\rho_\text{in}}^{\rho(N)} \frac{d\rho}{\partial_\rho(\ln X(\rho))} = -2N\ , \label{eqn:rhosolutionHJ}
\end{align}
where we put $N_\text{in}=0$ and we denote $\Delta N \equiv N_\text{end}$. We first solve the equation for the radius \autoref{eqn:rhosolutionHJ} to second order in $\delta\rho\equiv \rho-\rho_0$, using that $X(\delta\rho)= 1+\frac{ \lambda  }{2}\delta\rho ^2+ \frac{\alpha }{6}\delta \rho ^3 $. This yields
\begin{equation}
\delta\rho(N) = e^{-2\lambda N}\left(\delta \rho_\text{in}  -\frac{ \alpha}{2\lambda}  \left(1-e^{-2\lambda N} \right)\delta \rho_\text{in}^2\right).
\label{eqn:HJsolutiondeltarho}
\end{equation}
Notice that the solution is well defined in the limit $\lambda \rightarrow 0$ by expanding the exponential. Alternatively, we arrive at the same solution if we perturb the full background equation of motion of $\rho$ around the Hamilton-Jacobi solution, and discard the faster decaying mode:
\begin{align}
\delta \ddot  \rho  + 3 & H  \delta \dot \rho +  \left(- 2 \left(\frac{f^{\prime}}{f^2}\right)^\prime H^2  + V_{\rho\rho}\right) \delta \rho \nonumber  \\
& + \frac{1}{2}\left(V_{\rho\rho\rho}- 2 \left(\frac{f^{\prime}}{f^2}H^2\right)^{\prime\prime} \right) \delta \rho^2 = 0
\label{eqn:BGrhoperturbed}
\end{align}
First of all, the bracket in front of $\delta \rho$ evaluates exactly to $\mu^2$, the mass of isocurvature perturbations, which equals $\mu^2 = 2\lambda(3-2\lambda-\epsilon )H^2$. The linearized equation of motion in efolds therefore has Lyapunov exponents $\omega_\pm = \frac{1}{2}(-3 +\epsilon \pm (3- \epsilon - 4\lambda))$ (as long as $3 -\epsilon - 4\lambda  >0$). The least decaying mode is therefore indeed $\omega_+ = -2\lambda$. Furthermore, the bracket multiplying $\delta\rho^2$ evaluates to $\alpha(3-6\lambda) + \epsilon(-\alpha \kappa + 4\lambda)$, yielding exactly the same correction term as in equation \eqref{eqn:HJsolutiondeltarho}.
We can directly read off the power spectrum of isocurvature perturbations $\s \equiv \frac{\delta \rho}{\sqrt{2\epsilon}}$ from equation \eqref{eqn:HJsolutiondeltarho}, which reads
\begin{equation}
 P_\s = \frac{H^2}{8\pi^2\epsilon} e^{-4\lambda N}.
\end{equation}
Finding the solution to \autoref{eqn:thetasolutionHJ} involves a few more steps. We first expand
\begin{equation}
\frac{1}{f(\rho)} \approx \frac{1}{f(\rho_0)}\left(1-\frac{2\delta\rho}{\kappa}+\frac{ \left(\kappa ^2 \mathbb{R}+6\right)\delta \rho ^2}{2 \kappa ^2}\right)\ .
\end{equation}
This together with \autoref{eqn:HJsolutiondeltarho} allows us to integrate the right hand side of \autoref{eqn:thetasolutionHJ} to second order in $\delta \rho_
\text{in}$, resulting in a functional $\mathcal{F}(\kappa, \mathbb{R}, m^2, \alpha, \delta\rho_\text{in}, \Delta N)$. The left-hand side of \autoref{eqn:thetasolutionHJ}  depends on $\theta_\text{in}$. Moreover, it also depends on $\theta_\text{end}$, which on its turn depends on $\delta\rho_\text{in}$ through the relation $\epsilon(\theta_\text{end},\delta\rho_\text{end})=1$. This requires an explicit functional form for $W(\theta)$, but we neglect this contribution. It is typically suppressed because either $\delta\rho_\text{end}$ has vanished, or otherwise $\lambda$ is very small. %Nevertheless, we specialize further to $W(\theta) \sim \theta^p$ and take this contribution into account.  %We neglect this dependence on $\delta\rho_\text{in}$  in the rest of our computation, although one could take it into account if $W(\theta)$ is explicitly known\footnote{We have computed it explicitly for power law inflation $W \sim \theta^p$ and found a slow roll suppressed contribution}.

In the $\delta N$ formalism one computes how the number of efolds of an inflationary trajectory $\Delta N$ changes as a function of $\delta\rho_\text{in}$ and $\theta_\text{in}$. However, we are not required to solve for $\Delta N(\theta_\text{in}, \rho_\text{in})$ explicitly, because to compute the superhorizon evolution of $\R$, we only need to know its derivatives with respect to the initial perturbations
\begin{equation}
\R \approx \delta \Delta N = N_a \delta \phi^a +\frac{1}{2}N_{ab} \delta \phi^a \delta \phi^b +\ldots\ .
\end{equation}
Here we denoted $N_a \equiv \frac{\partial \Delta N}{\partial \phi^a}$ and $N_{ab} \equiv \frac{\partial^2 \Delta N}{\partial \phi^a \partial \phi^b}$, and $\delta\phi^a$ are the typical size of field fluctuations at horizon crossing.
Therefore, we proceed by taking derivatives of \autoref{eqn:thetasolutionHJ} with respect to $\theta_\text{in}$ and $\delta \rho_\text{in}$ which we can invert to find to linear order
\begin{align}
&\frac{N_\theta}{\sqrt{f_0}} = \sqrt{\frac{1}{2\epsilon}}, \\
&N_\rho = \frac{1}{\kappa\lambda }\left(1-e^{-2\lambda \Delta N} \right)\label{eqn:Nrho} ,
\end{align} 
where we have used that $\partial_\theta (\ln W(\theta)) = \sqrt{\frac{\epsilon f}{2}}$. This yields the power spectrum of curvature perturbations
\begin{equation}
P_\R = \left(\frac{H}{2\pi}\right)^2\left(\frac{N_\theta^2}{f}+N_\rho^2\right) = \frac{H^2}{8\pi^2 \epsilon}\left(1+\dd \right),
\label{eqn:deltaNpredictionPS}
\end{equation}
with $\dd\equiv 2\epsilon N_\rho^2$. We confirm this result in the next subsection by solving the perturbation equations explicitly. 

Similarly, we can compute the amplitude of the bispectrum generated on superhorizon scales. Assuming that the curvature perturbations are dominantly sourced by the radial isocurvature perturbations, we expect that the $\delta N$ formalism captures the bispectrum well. The Hessian $N_{ab}$ is given by
\begin{align}
&\frac{N_{\theta\theta}}{f_0} = -\frac{\eta}{4\epsilon}, \\
&\frac{N_{\theta\rho}}{\sqrt{f_0}} = \frac{ (1-\kappa \lambda N_\rho )}{\kappa}\sqrt{\frac{2}{\epsilon}} ,\\ 
& N_{\rho\rho} =-N_\rho^2 \left(\frac{\alpha \kappa}{2}+\lambda \frac{10-\mathbb{R} \kappa^2}{4}\right)+N_\rho\frac{2-\mathbb{R} \kappa^2}{2\kappa}.
\label{eqn:Nrhorho} 
\end{align}
%&+N_\rho\left(\frac{2-\mathbb{R} \kappa^2}{2\kappa} +\alpha p(2+5p\lambda)\right)
%& N_{\rho\rho} = -\frac{p \left(8 \kappa ^2 \lambda ^2-2 \alpha  \kappa  %p+\lambda  p \left(\kappa ^2 \mathbb{R}+6\right)\right)}{4 \kappa ^2 (1-%\lambda p)^2} \nonumber \\
%& + \frac{1- \lambda  p^2 (\frac{\alpha  \kappa}{2} -2 \lambda )+p \left(4 %\kappa ^2 \lambda ^3-2 \lambda \right)+\kappa ^2 \mathbb{R} ( \lambda 
%   p-\frac{1}{2})}{ \kappa  (1-\lambda  p)^2} N_\rho \nonumber \\
% & -  \frac{\lambda  \left(4 \lambda ^2 p^2+ \lambda  p \left(2 \kappa ^2 %\lambda ^2-5\right)+\frac{5}{2}\right)+(\alpha  \kappa-\frac{\kappa ^2 %\lambda  \mathbb{R}}{2}) (\frac{1}{2}- \lambda  p)}{ (1-\lambda  p)^2} %%\nonumber \\
% & \quad \times N_\rho^2 
%\begin{align}
%N_{\rho\rho} = -& \frac{1}{2}N_\rho^2 \left(\alpha \kappa(1+4p\lambda) +\lambda \frac{10-\mathbb{R} \kappa^2}{2}\right) \nonumber \\
%&+N_\rho\left(\frac{2-\mathbb{R} \kappa^2}{2\kappa} +\alpha p(2+5p\lambda)\right).
%\label{eqn:Nrhorho}
%\end{align}

The amplitude of local non-Gaussianities generated on superhorizon scales is given by \cite{Lyth:2005fi,Seery:2005gb} $f^{\delta N}_\text{NL} = -\frac{5}{6}\frac{G^{ab}G^{cd} N_a N_c \left(N_{bd}+\Gamma^g_{bd} N_b N_g\right)}{(G^{ef}N_e N_f)^2}$. 
Therefore, in the limit $\dd \gg 1$ we approximate the amplitude of the bispectrum to be $f_\text{NL} \approx \frac{5 N_{\rho\rho} }{6 N_\rho^2} $, that is,
\begin{equation}
f_\text{NL} \approx -\frac{5}{12} \left(\alpha \kappa +\lambda \frac{10-\mathbb{R} \kappa^2}{2}\right)+\frac{5}{N_\rho}\frac{2-\mathbb{R}\kappa^2}{12\kappa} \ .
\end{equation}
In particular, taking $\lambda = 0$ and $\alpha =0 $ we recover the results from \cite{Achucarro:2019pux}. A non-zero $\lambda$ or cubic coupling $\alpha$ will substantially enhance the amount of non-Gaussianities, though. Moreover, note that for small values of $\lambda$ and $\alpha \sim \frac{2-\mathbb{R}\kappa^2}{\kappa^2 N_\rho}$ the first and second term in $f_{NL}$ approximately cancel each other, and the other corrections to $f_{NL}$ should be taken into account.

Summarizing, in the limit $\dd \gg 1$ we find the following predictions for the tensor-to-scalar ratio, the spectral index, the tensor tilt, the amplitude of the reduced bispectrum, and its tilt, respectively:  
 \begin{subequations}
\begin{align}
  &r = \frac{16\epsilon}{1+\dd},\\
%  &n_s = 1 - 2\epsilon - 4\lambda e^{-2\lambda  \Delta N}\left(1-%e^{-2\lambda  \Delta N}\right)^{-1},\\
& n_s = 1 -2 \epsilon - \frac{1-4\kappa \lambda N_\rho}{N_\rho}, \\
  &n_t = -2\epsilon,\\
  &f_\text{NL} = -\frac{5}{12} \left(\alpha \kappa +\lambda \frac{10-\mathbb{R} \kappa^2}{2}\right)+\frac{5}{N_\rho}\frac{2-\mathbb{R} \kappa^2}{12\kappa},\\
 & n_{f_\text{NL}} \equiv \frac{d f_\text{NL}/ d\Delta N}{f_\text{NL}} = \frac{2-\mathbb{R} \kappa^2}{\kappa^2}\frac{1-\kappa \lambda N_\rho}{N_{\rho\rho}} \ ,
\end{align}
\label{eqn:predictionobservables}
\end{subequations}
with $\dd = 2\epsilon N_\rho^2$, and $N_\rho$ and $N_{\rho\rho}$ are given in \autoref{eqn:Nrho} and \autoref{eqn:Nrhorho}. 

We check these expressions numerically using \textit{PyTransport} \cite{Mulryne:2016mzv, Ronayne:2017qzn} (see also \cite{Dias:2016rjq, Seery:2016lko}). We take $\lambda \in \tfrac{1}{6}[0, 0.2 ]$ in six steps, $\alpha \in \{0.01,1,100\}$ and use polar coordinates $f(\rho) = \rho^2$ to inflate at a radius $\rho_0=2$. Smaller values of $\rho_0 \lesssim 1$ lead to numerical instabilities. In the inflationary direction we choose chaotic inflation with $W(\theta)\sim \theta$. First of all, we find that the analytical prediction for the amplitude of the power spectrum in \autoref{eqn:deltaNpredictionPS} is recovered within $1.0\%$ to $3.4\%$ precision in the range $\Delta N \in [50,60]$. The analytical prediction for $n_s$ on the other hand are accurate up to $0.03 \%- 0.31 \%$. In fact, the prediction for the amplitude is systematically a bit too small. This could for instance be due to corrections to the horizon crossing formalism. 
To check our predictions for $f_{NL}$ we take a squeezing ratio of $k_S/k_L = 9.5$ and let the sum of wavenumbers $K \equiv 2 k_S + k_L$ cross the horizon between 50 and 60 e-folds before the end of inflation. For small values of $\lambda$ and $\alpha = 0.01$ we find that the tilt of $f_{NL}$ is captured quite well with the analytical estimate for $n_{f_{NL}}$ above. The amplitude of $f_{NL}$, however, crosses zero for this value of $\alpha$, and is therefore sensitive to the remaining $\delta N$ corrections. We indeed find that including these corrections substantially improves agreement with the full numerical result. For larger values of $\alpha$ the $\delta N$ corrections become less important for the bispectrum amplitude, but more relevant for estimating its tilt. Finally, since we choose $\rho_0=2$ we break the condition $\dd \gg 1$ for moderately small values of $\lambda$ already. Numerically we confirm that, for increasing $\lambda$, the $\delta N$ corrections become important quickly. Nevertheless, the full $\delta N$ result captures both the amplitude and the spectral tilt of the squeezed bispectrum well.  

We conclude that the simple analytical predictions work well provided that 1) $\dd \gg 1$ and 2) $ \alpha \lesssim \O(1)$, but excluding a small regime around $\alpha \sim \frac{2-\mathbb{R}\kappa^2}{\kappa^2 N_\rho}$. Somewhat smaller values of $\dd$ and larger values of $\alpha$ yield a squeezed bispectrum that is better captured by the full $\delta N$ expression, as long as $\lambda$ is sufficiently small for the bispectrum to be dominantly sourced by superhorizon evolution.

\subsection{Solving the system of coupled perturbations}
\label{sec:coupledperturbations}
To confirm the $\delta N$ prediction of the power spectrum we solve the system of coupled perturbations directly.
Our starting point is the quadratic action for field perturbations \cite{Seery:2005gb, Langlois:2010vx, Achucarro:2010da, Gong:2011uw, Elliston:2012ab}
\cite{Achucarro:2016fby}
\begin{equation}
\begin{aligned}
 S^{(2)} = \frac{1}{2}\int d^4 x a^3&\left[ 2\epsilon\left({\dot\R}-\frac{2 H}{\kappa} \sigma\right)^2 - 2\epsilon \frac{(\partial_i \R)^2}{a^2}  \right. \\
 &  \ \quad \qquad \left. + {\dot\sigma}^2 -\mu^2 \sigma^2 -\frac{(\partial_i \sigma)^2}{a^2}  \right]
 \end{aligned}
  \label{eqn:quadraticactionperturbationsorbitalinflation}
\end{equation}
% \begin{widetext}
% \begin{equation}
%  S^{(2)} = \frac{1}{2} \int d^4 x a^3M_p^2 \left[2\epsilon\left({\dot\R}-\frac{2M_p H}{\kappa} \sigma\right)^2+{\dot\sigma}^2 -\mu^2 \sigma^2 -\frac{(\partial_i \sigma)^2}{a^2} - 2\epsilon \frac{(\partial_i \R)^2}{a^2} \right]\ .
%  \label{eqn:quadraticactionperturbationsorbitalinflation}
% \end{equation}
% \end{widetext}
Here $\R$ denotes the curvature perturbation, which is the only degree of freedom in single field inflation. Moreover, the isocurvature perturbations $\sigma \equiv (N)_a Q^a$ represent perturbations normal to the inflationary trajectory \cite{Gordon:2000hv}. Here $Q^a$ are the gauge invariant field fluctuations, equal to $\delta \phi^a$ in the flat gauge. If we put the $\sigma$-terms to zero the action reduces to the familiar quadratic action of single field inflation. \\

We solve for the superhorizon evolution of the curvature and isocurvature perturbations. Neglecting decaying solutions, their equations of motion simplify to 
\begin{subequations}
\begin{empheq}{align}
 &\R^\prime - \frac{2}{\kappa}\sigma = 0, \label{eqn:superhorizonR} \\
 & {\sigma}^{\prime\prime} +(3-\epsilon){\sigma}^\prime + \frac{\mu^2}{H^2}\sigma  = 0. \label{eqn:superhorizonS}
\end{empheq}
 \label{eqn:eomPTsh}
\end{subequations}
Here a prime denotes a derivative with respect to efolds $dN = H dt$.
For constant $0<\mu^2/H^2 = 2\lambda(3-2\lambda-\epsilon ) \ll 9/4$ and $\epsilon \ll 1$, the equation for isocurvature perturbations describes an overdamped oscillator with the same Lyapunov exponents as given below \eqref{eqn:BGrhoperturbed}, with $\omega= - 2\lambda$ as least decaying mode.  
The isocurvature perturbations in turn source the curvature perturbations according to \autoref{eqn:superhorizonR}. Integrating this equation for constant $\kappa$ gives the superhorizon solution for $\R$. 
 \begin{equation}
  \R(N) \approx \R_0 +\frac{\sigma_0 }{\lambda \kappa}\left(1-\exp\left( 2 \lambda  N\right)\right)\ .
  \label{eqn:solutioncurvatureorbitalinflation}
 \end{equation}
 Here we made use of the fact that $\kappa$ is constant. In the limit that $\mu^2/H^2=0$, the second term in \autoref{eqn:solutioncurvatureorbitalinflation} becomes proportional to $\Delta N/\kappa$.
In the quantum analysis \cite{GrootNibbelink:2000vx} there are two uncorrelated contributions to $\hat \R$. The first contribution is sourced by initial curvature perturbations where $\sigma_0 = 0$. This corresponds to a constant mode $\R_0$ that freezes out on super-Hubble scales. The second contribution is sourced by initial isocurvature perturbations where $\R_0 = 0$ and grows on superhorizon scales. Using the typical amplitude of quantum perturbations at horizon crossing $\sigma_0 \sim \sqrt{2\epsilon}\R_0 \sim \frac{H}{2\pi}$, we arrive at our simple estimate of the power spectrum of curvature perturbations
\begin{equation}
 P_\R \approx \frac{H^2}{8\pi^2 \epsilon}\left(1+ \frac{2\epsilon}{\lambda^2 \kappa^2}\left(1-e^{-2\lambda\Delta N}  \right)^2\right)
 %\label{eqn:powerspectrumorbitalinflation}
\end{equation}
Here all variables are understood to be evaluated at horizon crossing, and $\Delta N$ denotes the number of efolds counted from when the observable modes cross the horizon until the end of inflation. %In the limit that $\mu^2/H^2\rightarrow 0$, the second term in \autoref{eqn:powerspectrumorbitalinflation} becomes proportional to $\Delta N^2/\kappa^2$. 
Similarly, for the isocurvature spectrum we find
\begin{equation}
P_\s \approx  \frac{H^2}{8\pi^2 \epsilon}e^{-4\lambda \Delta N},
\end{equation}
where we defined $\s\equiv \frac{1}{\sqrt{2\epsilon}} \sigma$. Both results are in agreement with the $\delta N$ estimates from the previous subsection.

%The analytical results are obtained by using 
%\begin{equation}
% n_s = \frac{\partial \ln P_\R}{\partial N}, \quad r = \frac{2H^2}{\pi^2M_p^2}\frac{1}{P_\R},
% \label{eqn:analyticalnsreps}\ .
%\end{equation}
%since the tensor spectrum remains unchanged with respect to the single field case. \\

\subsection{Perturbative constraints}
 To ensure the validity of the perturbative approach we implicitly assumed, we should take $\kappa$ large enough. In the same way, the numerical code \cite{Mulryne:2016mzv} is performing a tree level in-in computation, and higher order tree level (and loop) corrections should be small compared to the leading result. Our simple analytical result captures the super-Hubble evolution of $\R$, and therefore provides an estimate of the leading order tree level computation.  Using \autoref{eqn:quadraticactionperturbationsorbitalinflation} we can estimate that $\xi \sim \frac{\sqrt{8\epsilon}}{\kappa}$ is the perturbation parameter that measures the relative size of the higher order tree level corrections compared to the leading tree level computation. In the in-in computation the sourcing of $\R$ by $\sigma$ is captured by the interaction term $S^{(2)}_\text{int} = \int d\tau d^3 x a^3 4\epsilon \frac{H}{\kappa}(\partial_\tau \R) \sigma $ (using conformal time $d\tau = dN/aH $.) Rewriting this in canonical variables $u\equiv \sqrt{2\epsilon}a \R$ and $v\equiv a \sigma$ we get $S^{(2)}_\text{int} \sim \int d\ln\tau d^3 x\ \xi (\partial_\tau u) v $, with $\xi = \frac{\sqrt{8\epsilon}}{\kappa}$. Therefore, we need to ensure that $\xi \ll 1$. This means we are save in \autoref{sec:phenomenology}, since we take $\kappa \geq 1$.\\

Moreover, we need to ensure that quantum perturbations remain much smaller than the radius of curvature $\delta \rho \ll \kappa$
In particular, we should be careful in the limit that the isocurvature perturbations are very light $\mu^2/H^2 \ll 1$. If we consider small values of $\kappa$, such that the second term in \autoref{eqn:powerspectrumorbitalinflation} dominates, we get $P_\R \sim \frac{H^2}{\kappa^2}$. Since the amplitude of the power spectrum is fixed by observations $A_\R \sim 10^{-9}$, this implies that the typical size of quantum fluctuations gets suppressed if we decrease $\kappa$. We find $\delta \rho^2 \sim H^2 \sim \kappa^2 A_\R \ll \kappa^2$, so we are always fine.

%Finally, we estimate the relative amplitude of isocurvature perturbations compared to the amplitude of curvature perturbations. How they are related to the late-time non-adiabatic perturbations depends on the mechanism of reheating, but we may still want to ensure that they are suppressed at the end of inflation. From \autoref{eqn:eomPTsh} it follows that the isocurvature perturbations decay on super-Hubble scales if $\mu^2/H^2 \gtrsim 3/\Delta N$, so we can assume that isocurvature perturbations with entropy masses of $\mu^2/H^2 \gtrsim 0.1$ have decayed by the end of inflation. However, for $\mu^2/H^2 \lesssim 0.1$, the ratio between curvature and isocurvature perturbations is given by 
%\begin{equation}
%\beta_\text{iso} \equiv \frac{P_\s}{ P_\R} \approx \frac{1}{1+8\epsilon \frac{M_p^2 \Delta N^2}{\kappa^2}}\ .
%\end{equation}
%Therefore, this leads to the constraint
%$\kappa^2/M_p^2 \ll 8\epsilon\Delta N^2$.

\end{document}